\documentclass[epsf]{aa}

\usepackage{times}
\usepackage{graphics}

\begin{document}

\title{{\it Letter to the Editor}\\
{\it ORFEUS\,} II echelle spectra :
Absorption by H$_2$ in the LMC\thanks{Data obtained under the 
DARA guest observing program in the {\it ORFEUS\,} II Mission}} 

\author{K.S. de Boer\inst{1}
\and P. Richter\inst{1}
\and D.J. Bomans\inst{2,}\inst{5}
\and A. Heithausen\inst{3}
\and J. Koornneef\inst{4}
}

\institute{
Sternwarte, Univ. Bonn, Auf dem H\"ugel 71, D-53121 Bonn, Germany
\and
Astron. Deptm., Univ. of Illinois at Urbana-Champaign,
1002 West Green St., Urbana, IL 61801, U.S.A.
\and
Radioastronomisches Institut, Univ. Bonn, Auf dem H\"ugel 71, 
        D-53121 Bonn, Germany
\and
Kapteyn Institute, Postbus 800, NL-9700AV Groningen, the Netherlands
\and
now at Astronomisches Institut, Ruhr-Universit\"at Bochum, Postfach 102148, 
        D-44780 Bochum, Germany
}

\date{Received 1 July 1998; Accepted 10 Aug. 1998}

\thesaurus{03.19.2, 08.09.2, 09.13.2, 11.09.4, 11.13.1, 13.21.3} 

\offprints{deboer@astro.uni-bonn.de}

\maketitle

\markboth{K.S. de Boer et al., {\it ORFEUS\,} observed H$_2$ absorption in LMC}
        {K.S. de Boer et al., {\it ORFEUS\,} observed H$_2$ absorption in LMC}

\begin{abstract}

We report the first detection of H$_2$\ absorption profiles of LMC gas 
on the line of sight to star 3120 in the association LH\,10 
near the emission nebula N\,11B.
Transitions found in the Lyman band 
are used to derive a total column density 
$N($H$_2)= 6.6 \times 10^{18}$ cm$^{-2}$.
Excitation temperatures of $\le 50$\,K for levels $J \leq 1$ and 
of $\simeq 470$ K for levels $2 \leq J \leq 4$  
of H$_2$\ are derived. 
We conclude that moderate UV pumping influences the population 
even of the lowest rotational states in this LMC gas. 

\keywords{Space vehicles - ISM: molecules - Galaxies: ISM - 
        Magellanic Clouds: LMC - Stars: individual: LH 10:3120
        - Ultraviolet: ISM}

\end{abstract}

\section{Introduction}

The last instrument capable of high resolution spectroscopy in the 900-1200 
\AA\ range was the {\it Copernicus\,} satellite (Spitzer et~al.~ 1973), 
working in the 1970s. 
This part of the spectrum contains absorption transitions of the 
Lyman and Werner Bands of molecular hydrogen, H$_2$, 
and of transitions of O\,{\sc vi} and other species 
highly relevant for interstellar medium studies. 

The {\it ORFEUS\,} (Kr\"amer et~al.~ 1990) and {\it IMAPS\,} (Jenkins et~al.~ 1988) 
experiments on the {\it ASTRO-SPAS\,} space shuttle platform 
has provided access to the far UV spectral range in great detail. 
The {\it ORFEUS\,} telescope feeds two spectrographs. 
The Heidelberg-T\"ubingen echelle gives spectra 
from 912 to 1410 \AA\ with $\lambda/\Delta\lambda \leq 10^4$ 
which are recorded with a microchannel plate detector. 
Good S/N is achieved in exposure times of the order of 1 hr 
of hot objects with $V < 12$ mag (Kr\"amer et~al.~ 1990). 
The UCB spectrograph produces spectra over the range of 390 to 1220 \AA\ 
(Hurwitz et~al.~ 1998) with a resolution too low for detailed interstellar work. 
{\it IMAPS\,} has its own telescope, works at 
$\lambda/\Delta\lambda \sim 1.5\cdot 10^5$ between 970 and 1195 \AA, 
and is in overall sensitivity limited to galactic studies 
(see e.g.\, Jenkins \& Peimbert 1997).

Here we report on the detection at high spectral resolution of H$_2$\ 
in the spectrum of the LMC star LH\,10:3120.
The star, located in the association LH\,10 near the western edge of the LMC, 
is of spectral type O5.5Vf, has $V=12.80$ mag and $E(B-V)=0.17$ mag 
(Parker et~al.~ 1992). 
The star was selected because of its very early spectral type, 
modest extinction (too large extinction would make the far-UV undetectable), 
and its proximity to an area where the molecule CO has been found in 
emission (Cohen et~al.~ 1988; Israel et~al.~ 1993).

The presence of H$_2$\ in the LMC is known since Israel \& Koornneef (1988) 
detected the near-IR emission lines from 
radiatively excited H$_2$\ seen toward H\,{\sc ii}\ regions near hot stars. 
Measurements showed that H$_2$\ is abundantly available, 
both in the SMC (Koornneef \& Israel 1985) 
and in the LMC (Israel \& Koornneef 1991a, 1991b).
Clayton et~al.~ (1996) detected with {\it HUT\,} at 3 \AA\ resolution 
broad depressions in the far-UV spectrum of two LMC stars, 
which could be fitted with H$_2$\ absorptions due to 
$N$(H$_2$) $\simeq$ 1-2 $10^{20}$ cm$^{-2}$.  
Studies of H$_2$\ in the LMC and SMC are of importance because of the 
lower metal content of these galaxies compared to the Milky Way 
and the different gas to dust ratios (see Koornneef 1984).

\begin{figure*}
\resizebox{\hsize}{!}{\includegraphics{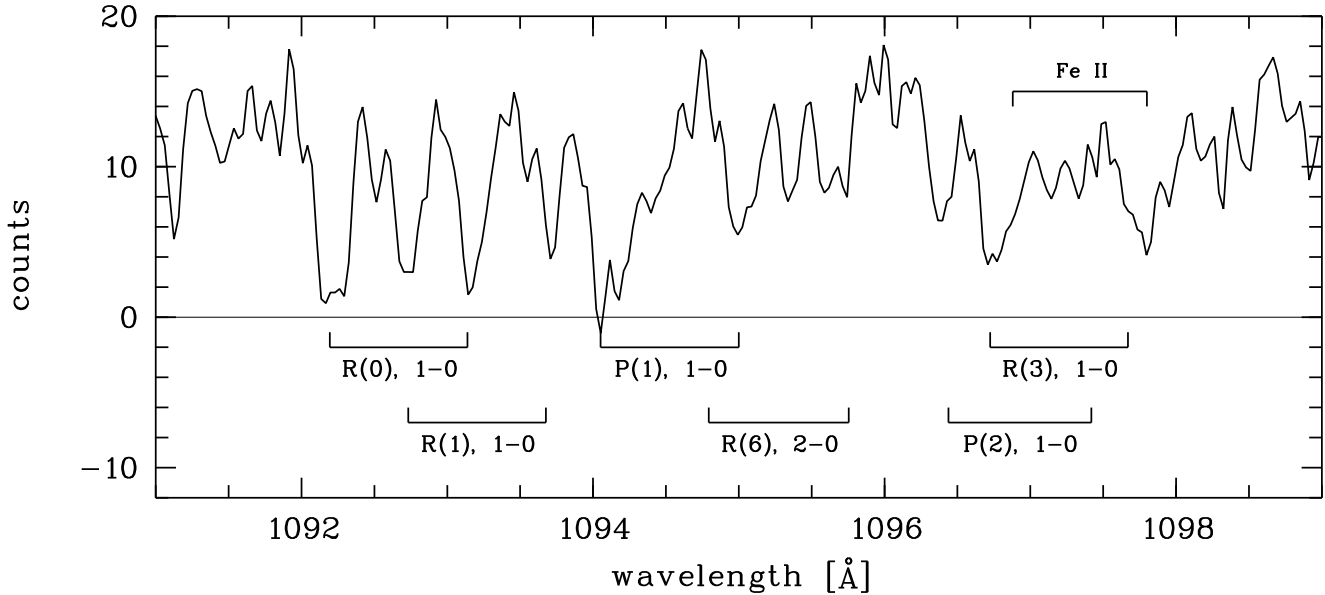}}
\caption[]{A portion of the {\it ORFEUS\,} spectrum of LH\,10:3120 
near 1095 \AA\ shows several H$_2$\ absorption lines. 
The strongest of these are identified by their transition below the spectrum.
The markings indicate the velocity range of 0 to 270 km\,s$^{-1}$\ 
for each of the lines.
Some weaker H$_2$\ lines as well as some atomic features
are also visible but have not been marked 
}
\end{figure*}

\begin{figure}
\resizebox{\hsize}{!}{\includegraphics{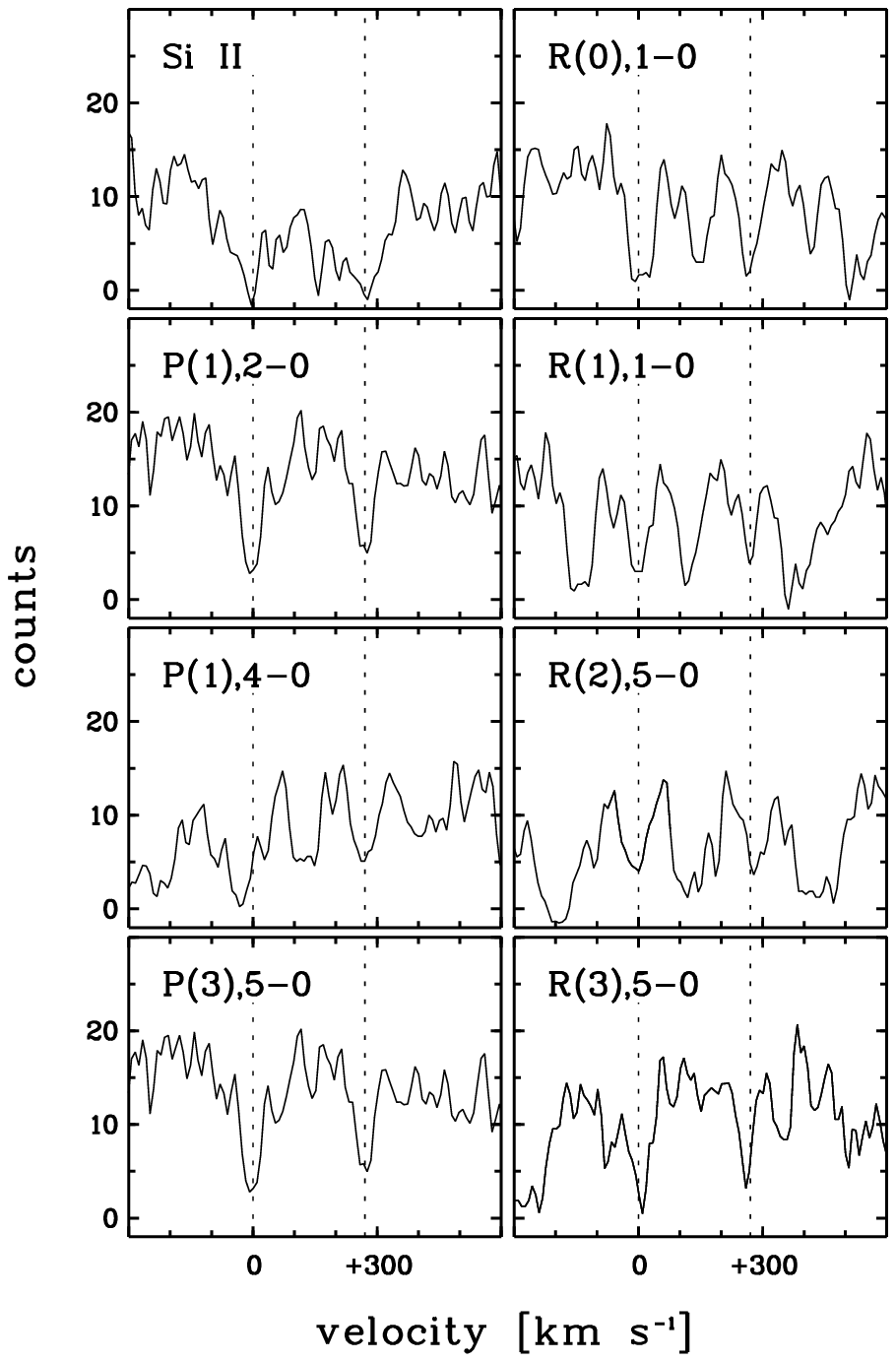}}
\caption[]{
Examples of absorption profiles clearly showing the detection of H$_2$.
At the top we present a profile of the Si\,{\sc ii} 1190.416 \AA\ line  
for easy reference. 
The H$_2$\ absorption is due to the Milky Way component near 0 km\,s$^{-1}$\
and the LMC component near +270 km\,s$^{-1}$.
The atomic line of Si\,{\sc ii} also shows the presence of the galactic halo 
high-velocity clouds near +60 and +130 km\,s$^{-1}$\ (LSR)
}
\end{figure}

\section{Observations, data handling}

The total observing time for LH\,10:3120\ in the {\it ORFEUS\,} space
shuttle mission of Nov./Dec. 1996 was 6000 s in 3 pointings exploiting 
the integrating capabilities of the microchannel plate detector system. 
A detailed instrument description and information about the 
basic data reduction is given by Barnstedt et al. (1998). 
The data reduction for the 20 echelle orders has been performed by 
the {\it ORFEUS\,} team in T\"ubingen.
The spectrum has been filtered by us with a de-noising algorithm 
basing on a wavelet transformation (Fligge \& Solanki 1997).
This leads to a slight degradation of the spectral resolution, 
now being equivalent to $~\simeq 30$ km\,s$^{-1}$. 
The spectra have a signal-to-noise (S/N) of $\simeq 20$ 
at the longer wavelengths of the recorded 
spectral range. 
Toward the shorter end of the spectral range, 
both the increased effect of the UV-extinction 
as well as the wavelength dependent sensitivity of the instrument 
leads to a reduction in S/N such, 
that little can be done with the spectrum at $\lambda < 1000$ \AA. 
After the filtering absorption features in the longer part of 
the spectrum become clearly visible.

\begin{table}[t]
\caption[]{H$_2$\ column densities in LMC gas toward LH\,10:3120\ }
\begin{tabular}{llcc}
\hline\noalign{\smallskip}
Rotation  & $\log N(J)$ & $b$-value & Number of \\
level $J$ & [cm$^{-2}$] & [km\,s$^{-1}$]   & lines used\\
\noalign{\smallskip}
\hline\noalign{\smallskip}
0 & 18.65 & 5 & 2 \\
1 & 18.10 & 5 & 5 \\
2 & 17.70 & 5 & 3 \\
3 & 17.50 & 5 & 3 \\
4 & 16.70 & 5 & 3 \\ 
\noalign{\smallskip}
\hline
\end{tabular}
\end{table}

We first inspected spectral ranges almost devoid of atomic absorption lines 
and where only few H$_2$\ absorptions are expected. 
The reason for this is to avoid getting confused 
in the search for H$_2$\ by the complexity of the absorption line profiles 
on the line of sight to the LMC. 
Yet, the characteristic pattern of absorption 
by the Milky Way disk near 0 km\,s$^{-1}$, 
by the LMC near +270 km\,s$^{-1}$, 
and possibly by high-velocity clouds near +60 and +130 km\,s$^{-1}$\ 
all known from {\it IUE\,} 
(see Savage \& de Boer 1979, 1981; de Boer et~al.~ 1980) 
and {\it HST\,} (Bomans et~al.~ 1995) spectra, 
helps to identify the absorption structures. 

A section of the echelle order 51, where several 
H$_2$\ aborption lines have been found, is shown in Fig.\,1. 
Many of the H$_2$\ profiles overlap in their $\sim$300 km\,s$^{-1}$\  wide 
profile structure and decompositions are not always possible.
However, in some cases the galactic absorption stayed unblended,
in other cases the LMC portion was blend free.
For this first analysis we took 16 H$_2$\ absorption lines 
from the lowest 5 rotational states for the further analysis. 
These lines are essentially free from any blending problems 
so that wrong identifications can be excluded. 
Absorption strengths could thus be determined for the LMC components
seen in these absorption profiles. 
A selection of characteristic H$_2$\ absorption line profiles in 
velocity scale (LSR) is shown in Fig.\,2.

\section{H$_2$\ column density in the LMC gas}

The absorption profiles have been analysed and decomposed into
the various velocity components.
Here we limit ourselves to the absorption pertaining to the LMC.
For each line the absorption equivalent width has been determined.
The $f$-values for the further analysis have been taken 
from Morton \& Dinerstein (1976) for the H$_2$\ transitions and for
the atomic absorption lines from the compilation of Morton (1991).
Theoretical Voigt profiles were
fitted to some of the identified H$_2$\ absorption lines and we thus
could derive an upper limit for the velocity dispersion
of $b<10$ km\,s$^{-1}$\ for the LMC gas.

Subsequently, curves of growth have been constructed for each of 
the absorptions by the 5 rotational states for the LMC component. 
The logarithmic equivalent widths for each rotational level $J$ have been
shifted horizontally to give a fit to a theoretical single cloud 
curve of growth as a function of log $N(J)f\lambda$.
The best fit for all 5 rotational states was obtained with $b=5$ km\,s$^{-1}$\ 
as shown in the empirical curve of growth for all identified H$_2$\ lines 
(Fig.\,3). 
The column densities $N(J)$ derived in this way have been collected in Table 1.
The uncertainties in the column densities are based on the quality of
the fits to the curve of growths.
They range from 0.2 to 0.4 in the logarithm, as shown in Fig.\,4.
The total column density for the lowest 5 rotational states 
is $N$(H$_2$)$_{\rm total}=6.6 \times 10^{18}$ cm$^{-2}$.

\section{Interpretation}

\subsection{Excitation state}

To explain the column densities and the relative population of the
rotational states of the H$_2$\ molecule we have to determine
the mechanism which is responsible for the excitation of the molecular
gas on our line of sight.
Two processes dominate the population of the excited states,
collisional excitation and pumping by UV photons (Spitzer \& Zweibel 1974).
With the column densities for the individual rotational states
we were able to determine the excitation temperature for the LMC gas
for $J \leq 4$.
In Fig.\,4 the LMC column densities $N(J)$ divided by their 
statistical weight $g_J$ are plotted against the excitation energy $E_J$. 
The equivalent excitation temperature of the gas can be derived by fitting
theoretical population densites from a Boltzmann distribution. 
For $J \ge 2$ we derive an equivalent excitation temperature of 470 K.

\begin{figure}
\resizebox{\hsize}{!}{\includegraphics{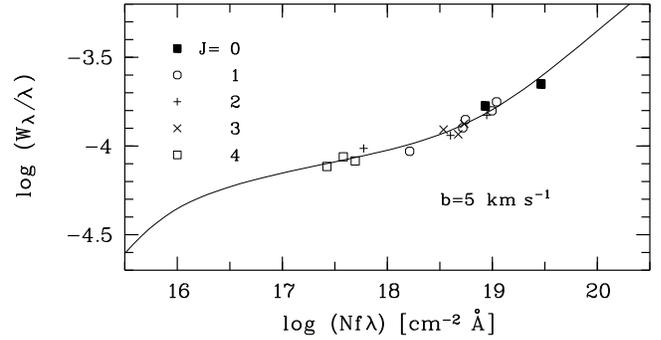}}
\caption[]{
The H$_2$\ lines were fitted to a single cloud curve of growth,
indicating $b \simeq 5$ km\,s$^{-1}$.
The column densities for the levels $J=0$ to 4 are given in Table 1
}
\end{figure}

\begin{figure}
\resizebox{\hsize}{!}{\includegraphics{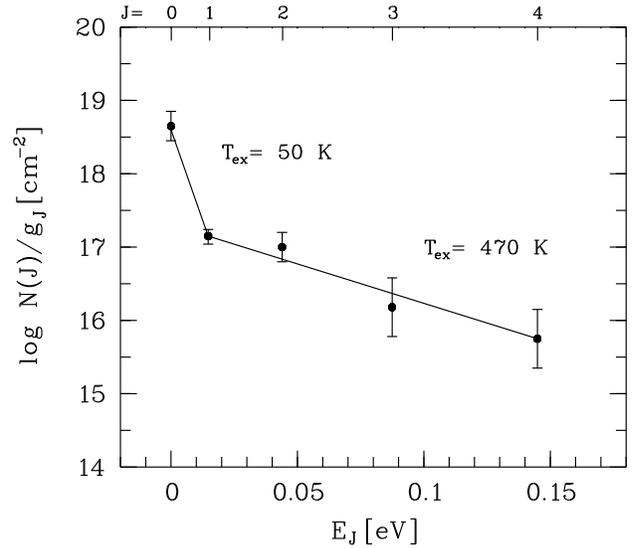}}
\caption[]{
The column densities for the levels $J=0$ to 4 of H$_2$\ are plotted against
the excitation energy.
The higher levels exhibit an equivalent excitation temperature of
$\simeq 470$ K, the excitation being mostly due to UV pumping.
For the two lowest levels the kinetic excitation temperature 
is $\le 50$ K; the fit drawn marks the indicated upper limit 
}
\end{figure}

The plot of Fig.\,4 shows also that the level $J=0$ lies above 
the linear relation of 470 K. 
Thus at least the $J=0$ level is populated through kinetic excitation. 
Fitting the Boltzmann relation to the $J = 0$ and 1 column densities 
we find a kinetic temperature of 50 K. 
Clearly, the population of level $J = 0$ is due to gas with 
a temperature $\le 50$ K. 
Such very low gas temperatures have also been found by Dickey et~al.~ (1994) 
and by Marx-Zimmer et~al.~ (1998) in the LMC through analysis of 
H\,{\sc i}\ 21 cm absorption. 

Israel \& Koornneef (1988) looked for emission by H$_2$\ 
in the direction of N\,11.
They report a marginal detection of H$_2$\ in the 1-0 $S(1)$ line 
at a position very close to our line of sight. 
However, given the difference in direction as well as the low significance 
of the emission detection a connection with our absorption 
will be speculative. 
Also, the emission may be from gas more distant than the star.

\subsection{Particularities of the line of sight}

Our line of sight through the LMC is ending at the star LH\,10:3120. 
The association LH\,10 ionizes the H\,{\sc ii}\ region N\,11B, 
the northern H\,{\sc ii}\ clump of the N\,11 superbubble complex.  
The central superbubble is apparently created by the association LH 9 
and is filled with hot gas.  
Several more patches of hot gas, 
partly coinciding with filamentary H$\alpha$ shells, exist.  
The one X-ray patch coinciding with N\,11B hints at wind driven bubbles 
around some massive stars inside N\,11B (Mac Low et~al.~ 1998).  
While N\,11B indeed shows first signs of the local effects 
of its most massive stars (Rosado et~al.~ 1996), it is still a 
relatively unevoled H\,{\sc ii}\ region without large-scale expansion. 
Our line of sight is therefore most likely illuminated by the stars of LH\,10. 
With the relatively unevolved nature of N\,11B it is tempting 
to relate the H$_2$\ absorbing gas with the remainder of 
the cold molecular cloud which formed N\,11B.

\subsection{Comparison with galactic gas}

The total column density found in the LMC can be related to the properties 
of the H$_2$\ gas in the Milky Way. 
The extinction to LH\,10:3120\ is $E(B-V)=0.17$. 
Taking out the galactic forground extinction of $E(B-V)=0.05$ 
(Parker et~al.~ 1992), 
one has extinction {\it in\,} the LMC of $E(B-V)=0.12$. 
The total H$_2$\ column density in the LMC is, 
compared to Milky Way gas (see Fig.\,4 of Savage et~al.~ 1977), 
slightly on the low side for its extinction.  
The smaller dust to gas content of the LMC 
(compared to the Milky Way; see Koornneef 1984) 
may have influenced the H$_2$\ to $E(B-V)$ ratio 
by way of a lower H$_2$\ formation rate. 

The excitation level of the higher $J$ levels on the line of sight 
to LH\,10:3120\ is equivalent to $\simeq 470$\,K. 
Such values are also found for the higher rotational states in galactic gas 
(Snow 1977; Spitzer et~al.~ 1974). 
The environment of N\,11B has contributed to the UV pumping 
but not in an excessive way. 
The excitation of the lowest level indicates the gas is kinetically cold.

\section{Concluding remarks}

Our investigation of the {\it ORFEUS\,} far UV spectrum 
shows for the first time well resolved absorption profiles due to H$_2$\ 
in LMC gas. 
The derived column densities and excitation temperatures toward LH 10:3120
demonstrate that UV pumping also takes place in LMC gas. 
The detected H$_2$\ has a column density and an excitation state 
similar to that of Milky Way gas, 
even though the line of sight runs through an energetic environment. 
In closing we note that H$_2$\ has also been detected in absorption 
by {\it ORFEUS\,} in the SMC (Richter et~al.~ 1998).

\acknowledgements
We thank the Heidelberg-T\"ubingen team for their support 
and the data reduction at the instrumental level. 
We thank in particular H. Widmann for helpful discussions.
{\it ORFEUS\,} could only be realized with the support of the German and
American colleagues and collaborators.  
The {\it ORFEUS\,} project was supported by DARA grant WE3 OS 8501, WE2 QV 9304 and
NASA grant NAG5-696. 
PR is supported by a grant from the DARA (now DLR) under code 50 QV 9701 3.

{}

\begin{thebibliography}{}

\bibitem[]{}
Barnstedt J., Kappelmann N., Appenzeller I., et al., 1998, A\&AS, subm.

\bibitem[]{}
Bomans D.J., de Boer K.S., Koornneef J., Grebel E.K., 1996, A\&A 313, 101

\bibitem[]{}
Clayton G.C., Green J., Wolff M.J., et al., 1996, ApJ 460, 313

\bibitem[]{}
Cohen R.S., Dame T.M., Garay G., Montani J., Rubio M., Thaddeus P., 
        1988, ApJ 331, L 95

\bibitem[]{}
de Boer K.S., Koornneef J., Savage B.D., 1980, ApJ 236, 769

\bibitem[]{}
Dickey J.M., Mebold U., Marx M., Amy S., Haynes R.F., Wilson W., 
        1994, A\&A 289, 357

\bibitem[]{}
Fligge M., Solanki S.K., 1997,  A\&AS 124, 579

\bibitem[]{}
Hurwitz M., Bowyer S., Bristol R., et al., 1998, ApJ 500, L 1

\bibitem[]{}
Israel F.P., Koornneef J., 1988, A\&A 190, 21

\bibitem[]{}
Israel F.P., Koornneef J., 1991a, A\&A 248, 404

\bibitem[]{}
Israel F.P., Koornneef J., 1991b, A\&A 250, 475

\bibitem[]{}
Israel F.P., Johansson L.E.B., Lequeux J., et al., 
        1993, A\&A 276, 25

\bibitem[]{}
Jenkins E.B., Peimbert A., 1997, ApJ 477, 265

\bibitem[]{}
Jenkins E.B., Joseph C.L., Long D., et al., 
        1988, Proc. SPIE, 932, p. 213

\bibitem[]{}
Koornneef J., 1984, in IAU Symp.108, `Structure and evolution of the 
        Magellanic Clouds', eds. S.\,van den Bergh \& K.S.\,de Boer; 
        Reidel, Dordrecht, p. 333

\bibitem[]{}
Koornneef J., Israel F.P., 1985, ApJ 291, 156

\bibitem[]{}
Kr\"amer G., Barnstedt J., Eberhard N., et al., 1990, in IAU Coll.\,123, 
        `Observatories in Earth Orbit and beyond', 
        ed.{} Y.\,Kondo, Kluwer, p. 177

\bibitem[]{}
Mac Low M.-M., Chang T.H., Chu Y.-H., Points S.D., Smith R.C., Wakker B.P., 
        1998, ApJ 493, 260

\bibitem[]{}
Marx-Zimmer M., Herbstmeier U., Dickey J.M., Zimmer F., Staveley-Smith L., 
Mebold U., 1998, A\&A, submitted


\bibitem[]{}
Morton D.C., 1991, ApJS 77, 119

\bibitem[]{}
Morton D.C., Dinerstein H.L., 1976, ApJ 204, 1

\bibitem[]{}
Parker J.W., Garmany C.D., Massey P., Walborn N.R., 1992, AJ 103, 1205

\bibitem[]{}
Richter P., Widmann H., de Boer K.S., et al., 1998, A\&A, Letter, subm.

\bibitem[]{}
Rosado M., Laval A., Le Coarer E., Georgelin Y.P., Amram P., Marcelin M., 
Goldes G., Gach J.L., 1996, A\&A 308, 588 

\bibitem[]{}
Savage B.D., de Boer K.S., 1979, ApJ 230, L 77

\bibitem[]{}
Savage B.D., de Boer K.S., 1981, ApJ 243, 460

\bibitem[]{}
Savage B.D., Bohlin R.C., Drake J.F., Budich W., 1977, ApJ 216, 291

\bibitem[]{}
Snow T.P., 1977, ApJ 216, 724

\bibitem[]{}
Spitzer L., Zweibel E.G., 1974, ApJ 191, L 127

\bibitem[]{}
Spitzer L., Drake J.F., Jenkins E.B., Morton D.C., Rogerson J.B., York D.G., 
        1973, ApJ 181, L 122

\bibitem[]{}
Spitzer L., Cochran W.D., Hirschfeld A., 1974, ApJS 28, 373

\bibitem[]{}

\end{thebibliography}
\end{document}